\batchmode
\makeatletter
\def\input@path{{C:/Users/cheng/Documents/Papers/Chgwan/25-WestD0/NF-108467R1/}}
\makeatother
\documentclass[twoside,english,sort&compress]{iopart}
\PassOptionsToPackage{natbib=true}{biblatex}
\usepackage[T1]{fontenc}
\usepackage{textcomp}
\usepackage[latin9]{inputenc}
\usepackage{xcolor}
\usepackage{babel}
\usepackage{array}
\usepackage{float}
\usepackage{booktabs}
\usepackage{multirow}
\usepackage{amstext}
\usepackage{graphicx}
\usepackage{geometry}
\PassOptionsToPackage{normalem}{ulem}
\usepackage{ulem}
\usepackage[pdftex,
 bookmarks=true,bookmarksnumbered=false,bookmarksopen=false,
 breaklinks=false,pdfborder={0 0 1},backref=false,colorlinks=true]
 {hyperref}
\hypersetup{
 pdfauthor={Mr. Wan},
  citecolor=blue,linkcolor=blue,urlcolor=blue}

\makeatletter

\providecommand{\tabularnewline}{\\}
\providecolor{lyxadded}{rgb}{0,0,1}
\providecolor{lyxdeleted}{rgb}{1,0,0}
\DeclareRobustCommand{\mklyxadded}[1]{\textcolor{lyxadded}\bgroup#1\egroup}
\DeclareRobustCommand{\mklyxdeleted}[1]{\textcolor{lyxdeleted}\bgroup\mklyxsout{#1}\egroup}
\DeclareRobustCommand{\mklyxsout}[1]{\ifx\\#1\else\sout{#1}\fi}

\usepackage{iopams}
\usepackage{setstack}

\AtBeginDocument{\setkeys{Gin}{width=0.125\textwidth}}

\newcommand{\eqref}[1]{(\ref{#1})}
\usepackage{fancyhdr}
\usepackage[justification=raggedright,labelfont=bf,singlelinecheck=true]{caption}
\usepackage[pagewise]{lineno}
\usepackage{tablefootnote}
\usepackage{placeins}

\makeatother

\usepackage[style=numeric,style=numeric-comp,sorting=none,natbib=true,maxnames=2,minnames=1,url=false,eprint=false]{biblatex}
\begin{document}
\title{Machine learning prediction of plasma behavior from discharge configurations
on WEST}
\author{Chenguang Wan$^{1,3}$, Feda Almuhisen$^{2,*}$, Philippe Moreau$^{2}$,
R\'emy Nouailletas$^{2}$, Zhisong Qu$^{1,*}$, Youngwoo Cho$^{1}$,
Robin Varennes$^{1}$, Kyungtak Lim$^{1}$, Kunpeng Li$^{1}$, Jia
Huang$^{3}$, Weidong Chen$^{4}$, Jiangang Li$^{3}$, and Xavier
Garbet$^{1,2}$}
\address{1. School of Physical and Mathematical Sciences, Nanyang Technological
University, Singapore 637371, Singapore}
\address{2. CEA, IRFM, F-13108 Saint Paul-lez-Durance, France}
\address{3. Institute of Plasma Physics, Hefei Institutes of Physical Science,
Chinese Academy of Sciences, Hefei 230031, China}
\address{4. School of Information Science and Technology, University of Science
and Technology of China, Hefei 230093, China}
\ead{\href{mailto:feda.almuhisen@cea.fr}{feda.almuhisen@cea.fr} and \href{mailto:zhisong.qu@ntu.edu.sg}{zhisong.qu@ntu.edu.sg}}
\begin{abstract}
Accurately predicting plasma behavior based on discharge configurations
is essential for the safe and efficient operation of tokamak experiments.
While physics-based integrated modeling codes provide valuable insights,
their high computational cost limits their applicability for fast
scenario design and control optimization. In this study, we propose
a transformer-based machine learning model to predict key global plasma
parameters on the WEST tokamak, including the normalized beta ($\beta_{n}$),
toroidal beta ($\beta_{t}$), poloidal beta ($\beta_{p}$), plasma
stored energy ($W_{\mathrm{mhd}}$), safety factor at the magnetic
axis ($q_{0}$), and safety factor at the 95\% flux surface ($q_{95}$).
The model uses only signals that can be defined before the discharge,
such as magnetic coil currents, auxiliary heating power, plasma current
reference, and line-averaged plasma density. Trained on 550 discharges
from the WEST campaigns, the model demonstrates an average mean square
error (MSE) loss of 0.026, an average coefficient of determination
$R^{2}$ of 0.94, and achieves inference times on the order of 0.1
seconds. These results highlight the potential of data-driven surrogate
models for assisting in discharge planning, scenario evaluation, and
real-time control of tokamak plasmas.
\end{abstract}
\maketitle
\section{Introduction}

Any plasma discharge needs to be prepared in advance using a discharge
schedule editor, in which all tokamak settings are defined. In order
to optimize the tokamak operation, these settings must be assessed
using a tokamak simulator in order to check the feasibility of the
preset discharge scenario.

Predicting plasma behavior from the inputs taken from the discharge
schedule editor in tokamaks has traditionally relied on first-principles
physics codes, commonly referred to as \textquotedbl Integrated Modeling\textquotedbl .
These include ETS \citep{Falchetto2014}, PTRANSP \citep{Budny2008},
TSC \citep{Kessel2006}, NICE \citep{Faugeras2020}, CRONOS \citep{Artaud2010},
JINTRAC \citep{Romanelli2014}, METIS \citep{Artaud2018}, ASTRA \citep{Pereverzew1991},
RAPTOR \citep{Felici2011}, and TOPICS \citep{Hayashi2010}, etc.
While these codes simulate key physical processes with high fidelity,
their substantial computational cost makes them unsuitable for fast
or real-time prediction tasks.

This is especially true when we need to quickly predict zero-dimensional
parameters, such as plasma stored energy $W_{mhd}$, toroidal beta
$\beta_{n}$, safety factor q at the magnetic axis $q_{0}$, from
a given set of coil currents and preset parameters. Moreover, in practical
applications, integrated models involve approximations, model coupling
assumptions, and parameter uncertainties that constrain their achievable
accuracy. Some real-time-oriented frameworks, such as RAPTOR \citep{VanMulders2026,Felici2011},
partially address computational efficiency, but rely heavily on reduced
physics models. Their predictive accuracy depends on empirical coefficients,
including transport scalings and source models, as well as careful
tuning, scenario-specific calibration, and expert knowledge of the
device and discharge conditions. While such approaches offer strong
physics consistency, control robustness, and interpretability, they
often sacrifice shot-by-shot fidelity and have limited ability to
capture hidden correlations present in large experimental datasets.

Data-driven methods offer a way to bridge this gap by learning the
input-output mapping directly from experimental data, thereby bypassing
the need to solve first-principles equations repeatedly. Because of
this advantage, researchers in the fusion community have widely adopted
data-driven approaches to study a range of problems. Early applications
focused on disruption prediction \citep{Rea2018,Kates-Harbeck2019,Zheng2023,Zheng2018}.
More recent studies have further advanced this area: the first domain-adaptation
framework for cross-tokamak disruption prediction achieved clear gains
\citep{Shen2024}, and physics-guided feature extraction has shown
clear benefits when data are limited while also improving model interpretability
\citep{Shen2023}. Additional applications include magnetic field
control \citep{Degrave2022}, a surrogate model for the Ion Temperature
Gradient (ITG) mode \citep{Wan2025}, surrogate model of QualiKiz
\citep{VanDePlassche2020}, tearing mode control \citep{Seo2024},
the first real-time avoidance of density-limit disruptions in J-TEXT
\citep{Zheng2018}, the last-closed flux surface reconstruction \citep{Wan2023,Wan2024},
electron temperature profile reconstructions \citep{Clayton2013},
equilibrium reconstruction \citep{Joung2020,Coccorese1994}, equilibrium
solver \citep{VanMilligen1995}, missing electron temperature reconstruction
\citep{Wang2025c}, density profile inversion \citep{Huang2025},
radiation power estimation \citep{Barana2002}, confinement time prediction
with uncertainty quantification \citep{Gao2025}, instability identification
\citep{Murari2013}, neutral beam effect estimation \citep{Boyer2019},
confinement state classification \citep{Matos2021}, scaling law determination
\citep{Murari2010,Gaudio2014}, density limit analysis \citep{Maris2025},
plasma detachment prediction \citep{Yu2025}, imaging neutral particle
analyzer \citep{Garcia2025}, the direct control of tokamak parameters
\citep{Degrave2022,Seo2024}, and plasma magnetic measurement estimation
\citep{Ling2025}.

Furthermore, some data-driven approaches have explored full discharge
prediction \citep{Wan2022,Wan2021,Char2024}. These models do not
rely solely on configurations but also incorporate signals such as
gas puffing system data, which are hard to determine prior to a discharge
and are influenced by the evolving plasma state during the discharge.
This dependence constrains their generalizability and practical applicability.

In this study, we developed a transformer-based machine learning model
that utilizes only signals that can be directly controlled or feedback-regulated.
Compared to previous works \citep{Wan2022,Wan2021}, our model does
not rely on the state of any specific discharge process, thereby enhancing
its practical applicability. Specifically, we successfully reproduced
six key zero-dimensional global parameters on the Tungsten (W) Environment
in Steady-State Tokamak (WEST) by inputting 19 signals, including
poloidal field coil currents, auxiliary heating, plasma reference
current, and plasma electron density measurements at the magnetic
axis.

Our model achieves an average MSE of 0.026, an average coefficient
of determination $R^{2}$ of 0.94, and an inference time of $\sim0.1$
seconds, enabling large-scale control parameter sweeps and optimization
tasks. The dataset comprises 550 stable and reproducible discharges
from WEST campaigns, all conducted under well-controlled conditions
with good wall conditions. This consistency ensures the robustness
and applicability of the model in similarly controlled experimental
settings.

The remainder of this paper is organized as follows. Section \ref{sec:Dataset}
describes the dataset preparation and the analysis of the input and
output signals. Section \ref{sec:Model} outlines the machine learning
methodology. Section \ref{sec:Results} presents the model results
along with the corresponding analysis. Finally, Section \ref{sec:Conclusion-and-discussion}
presents a brief conclusion and future directions.

\section{Dataset \label{sec:Dataset}}

The WEST tokamak is equipped with a comprehensive suite of diagnostic
systems, including magnetic diagnostics \citep{Moreau2018} and plasma
state diagnostics \citep{Bouchand2025,Meyer2018,Gil2019}, among others.
In the present study, as shown in table \ref{tab:Signal_list}, the
input signals include the power and corresponding phase of the Lower
Hybrid Wave (LHW) Current Drive and Heating System, the power of the
Ion Cyclotron Resonance Heating (ICRH) System, the reference plasma
current, the currents in the Poloidal Field (PF) coils, and the real-time
measured plasma line-average density. The output signals include normalized
beta $\beta_{n}$, toroidal beta $\beta_{t}$, poloidal beta $\beta_{p}$,
plasma stored energy $W_{mhd}$, safety factor $q$ at magnetic axis
$q_{0}$ and safety factor $q$ at the 95\% flux surface.

It should be noted that we use $n_{e}$\_real as a proxy that captures
the combined influence of the fueling strategy and wall conditions.
In WEST, the electron density is primarily controlled through the
fueling system \citep{Nouailletas2023}, e.g. the gas injection system.
However, this control is neither instantaneous nor one-to-one. The
fueling system exhibits a relatively long response time, and the achieved
density depends nonlinearly on wall conditions and turbulent transport.
The main objective of the present work is to predict plasma responses
from discharge design parameters, also referred to as experimental
proposals, in which the target density is assumed to be achievable
in the experiment through feedback control.

\begin{table}
\caption{The list of signals. A signal name ending with \textquotedbl\_real\textquotedbl{}
indicates the real diagnostic data during the tokamak discharge process.
The signal name ending with \textquotedbl\_ref\textquotedbl{} indicates
that the data is reference data, i.e. nominal signals in the Discharge
Control System (DCS) of WEST. \label{tab:Signal_list}}

\centering{}%
\begin{tabular}{>{\raggedright}m{0.1\linewidth}>{\centering}p{0.5\linewidth}>{\centering}p{0.1\linewidth}>{\centering}p{0.1\linewidth}}
\hline 
\multirow{1}{0.1\linewidth}{\raggedright{}Signals} &
\raggedright{}Physics meanings &
\# of channels &
sampling rate\tabularnewline
\hline 
\multicolumn{2}{l}{Output signals} &
6 &
\tabularnewline
\hline 
\raggedright{}$\beta_{n}$ &
\raggedright{}Normalized beta &
1 &
1 - 12 Hz\tabularnewline
\raggedright{}$\beta_{t}$ &
\raggedright{}Toroidal beta &
1 &
1 - 12 Hz\tabularnewline
\raggedright{}$\beta_{p}$ &
\raggedright{}Poloidal beta &
1 &
1 - 12 Hz\tabularnewline
\raggedright{}$W_{mhd}$ &
\raggedright{}Plasma stored energy &
1 &
1 - 12 Hz\tabularnewline
\raggedright{}$q_{95}$ &
\raggedright{}Safety factor q at the 95\% flux surface &
1 &
1 - 12 Hz\tabularnewline
\raggedright{}$q_{0}$ &
\raggedright{}Safety factor q at the magnetic axis &
1 &
1 - 12 Hz\tabularnewline
\hline 
\multicolumn{2}{l}{Input signals} &
19 &
\tabularnewline
\hline 
\raggedright{}LHW\_real &
\raggedright{}Power of Lower Hybrid Wave (LHW) Current Drive and Heating
System &
4 &
1 kHz\tabularnewline
\raggedright{}ICRH\_real &
\raggedright{}Power of Ion Cyclotron Resonance Heating (ICRH) System &
3 &
1 kHz\tabularnewline
\raggedright{}$I_{p}$\_ref &
\raggedright{}Reference of plasma current &
1 &
1 kHz\tabularnewline
\raggedright{}$n_{e}$\_real &
\raggedright{}Actual line-average electron density at the magnetic
axis &
1 &
1 kHz\tabularnewline
\raggedright{}PF\_real &
\raggedright{}Current of Poloidal Field (PF) coils &
10 &
1 kHz\tabularnewline
\hline 
\end{tabular}
\end{table}

A total of 550 discharges, ranging from discharge \#57381 to \#60286,
were initially selected from the WEST database. Discharges with durations
less than $2$ seconds were excluded, as the ramp-up phase in the
WEST tokamak typically takes about 2 seconds. Such short discharges
are usually indicative of abnormal events and were not included in
the analysis. All selected signals were uniformly resampled at 1 kHz
from the start of each discharge and aligned to a common time base.
The data were saved in HDF5 format on a discharge-by-discharge basis,
resulting in approximately 15 gigabytes of original data. Across the
full dataset, the total Lower Hybrid Wave (LHW) power ranges from
0 to $\sim4$ MW, while the total Ion Cyclotron Resonance Heating
(ICRH) power ranges from 0 to $\sim3.1$ MW. The plasma current $I_{p}$
spans from $\sim$100 to $\sim$700 kA, with most discharges concentrated
around 400, 500, and 700 kA. The electron density varies from $\sim1.0$
$\times10^{19}$ to $\sim6.2\times10^{19}$ m$^{-3}$. The normalized
beta $\beta_{n}$ ranges from 0.17 to 1.01, the poloidal beta $\beta_{p}$
from 0.33 to 1.57, and the toroidal beta $\beta_{t}$ from 0 to $3\times10^{-3}$.
The plasma stored energy $W_{mhd}$ ranges from 212 to 342 kJ. The
safety factor $q_{95}$ spans from 0.88 to 8.64, while the central
safety factor $q_{0}$ ranges from 0.96 to 3.25. The details of the
distribution are given in Appendix \hyperlink{AppendixA}{A} figure
\ref{fig:Output-dist}.

\begin{figure}
\begin{centering}
\includegraphics[width=0.5\textwidth]{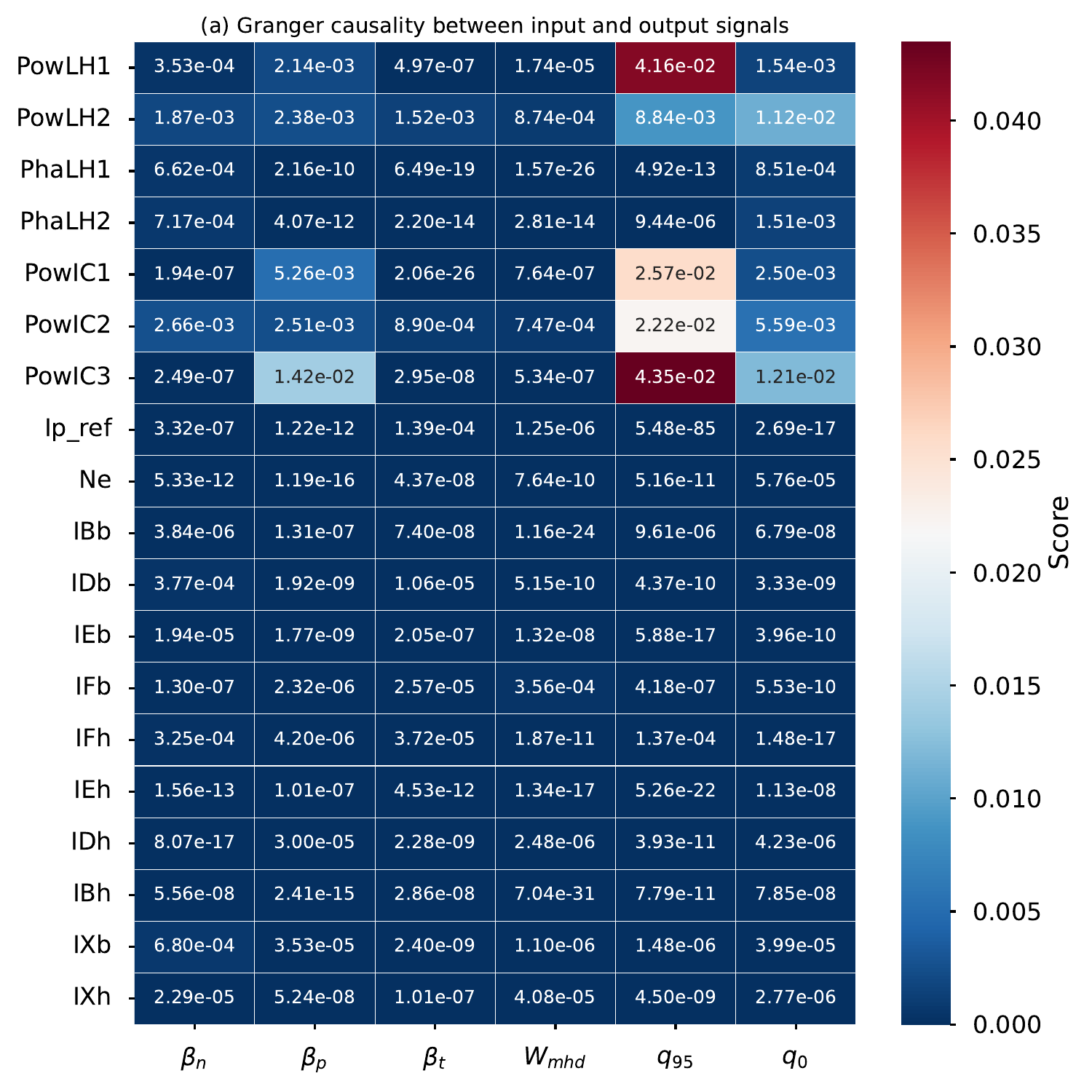}\includegraphics[width=0.5\textwidth]{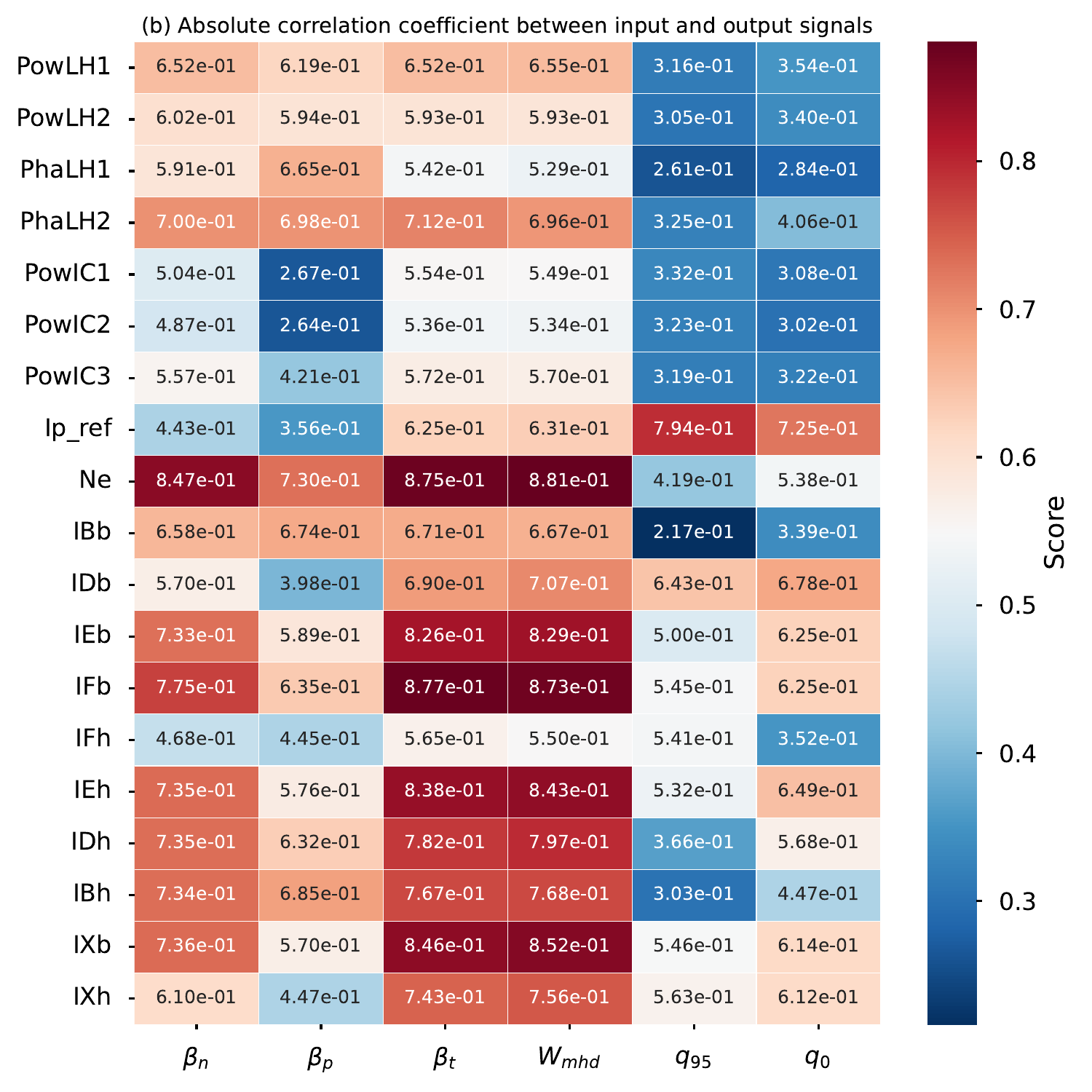}
\par\end{centering}
\caption{\label{fig:granger-pearson} The Granger causality and Pearson correlation
coefficient between input and output signals. For Granger causality,
a smaller coefficient indicates a stronger causal relationship, whereas
for the Pearson correlation coefficient, a larger value reflects a
stronger linear correlation. PowerLH{*} and PhaLH{*} are the actual
powers and corresponding phases of the LHW system, PowerIC{*} is the
powers of the ICRH system, Ip is the reference of plasma current,
the remaining signals correspond to various locations of the PF coils.}
\end{figure}

The selection of input signals was guided by the experience of operators
and plasma physicists. To further validate the relevance of these
inputs prior to training the machine learning model, and to assess
the relationships between input and output signals, we computed both
the Granger causality p-values and the Pearson correlation coefficients
($r$). As shown in figure \ref{fig:granger-pearson} (a), Granger
causality analysis reveals that all 19 input signals exhibit statistically
significant Granger causal relationships with the output signals,
with p-values below 0.05. Figure \ref{fig:granger-pearson} (b) shows
that the Pearson correlation coefficients between the input and output
signals are also statistically significant, indicating strong linear
associations.

For the Granger causality analysis, non-stationarity is common in
our signals. Therefore, we tested lags from 1 to 4 and employed four
different statistical methods: the SSR-based F test, the SSR-based
$\chi^{2}$ test, the likelihood ratio (LR) test, and the parameter
F test. This resulted in 4\texttimes 4 p-values, from which we selected
the minimum value as the final p-value. For the Pearson correlation
coefficient, we also introduced a lag term and focused solely on the
absolute magnitude of the correlation, disregarding the sign (positive
or negative). We tested lags from 0 to 4, and selected the maximum
absolute value as the final correlation coefficient $r$. The final
correlation coefficient $r_{\text{final}}$ was defined as:

\[
r_{\text{final}}=\max_{l\in\{0,1,2,3,4\}}\left|r\left(\{x_{l},x_{l+1},\dots,x_{n}\},\{y_{0},y_{1},\dots,y_{n-l}\}\right)\right|,
\]

where $r$ denotes the standard Pearson correlation coefficient, and
$l$ is the lag.

While these statistical tests offer valuable preliminary insights,
it is important to note that machine learning models possess strong
nonlinear mapping capabilities. Traditional statistical measures such
as Granger causality and Pearson correlation capture only specific
types of relationships and may not fully reveal the complex dependencies
within the data. Therefore, the statistical analysis primarily serves
to provide prior knowledge and validate the relevance of the input
features. Machine learning methods remain essential for constructing
the complex mappings between input and output signals.

\section{Methods \label{sec:Model}}

\begin{figure}
\begin{centering}
\includegraphics[width=1\textwidth]{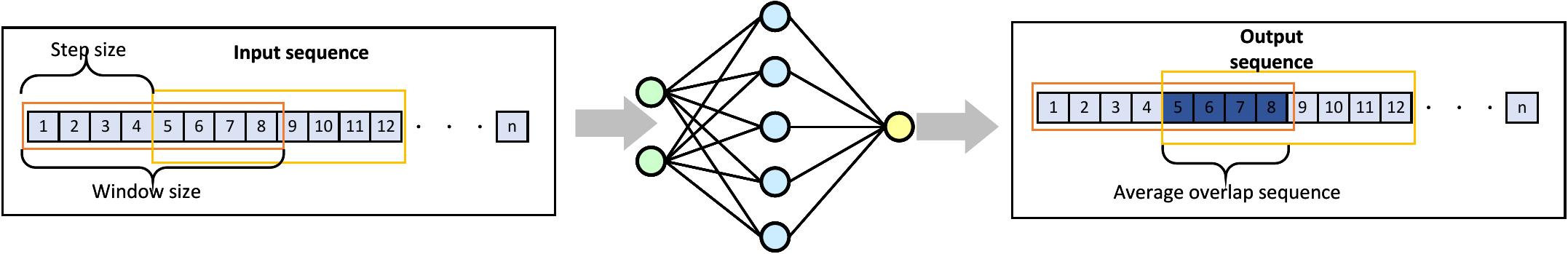}
\par\end{centering}
\caption{The workflow of machine learning model in the present work \label{fig:workflow}}
\end{figure}

The machine learning model and its data processing workflow are shown
in figure \ref{fig:workflow}. Prior to model input, the raw data
are smoothed using a Simple Moving Average (SMA) filter to suppress
noise. The SMA operation is defined in equation \ref{eq:1}. To accommodate
discharge sequences of varying lengths, the data are segmented using
a fixed-size sliding window with a window length of 1024 and a step
size of 512. This configuration yields a data augmentation ratio of
2, resulting in a total of $12059\times2$ sliced windows, where 2
denotes the augmentation factor introduced by overlapping segmentation.
Because the output sequences overlap as shown by the dark blue regions
in figure \ref{fig:workflow}, the overlapping segments are averaged
to improve numerical stability and prediction accuracy (see Appendix
\hyperlink{AppendixB}{B}, figure \ref{fig:Variance} for supporting
evidence). The 550 discharges were randomly divided into training
(60\%), validation (20\%), and test (20\%) sets. A strictly chronological
split was not used, as the objective of the present study is to evaluate
inter-discharge generalization within the operational envelope represented
in the WEST dataset, and to provide a fast surrogate for scenario
evaluation and discharge design under conditions similar to those
already explored. The SMA filter is defined as:

\begin{equation}
\mathrm{SMA}_{t}=\frac{1}{w}\sum_{i=-\lfloor w/2\rfloor}^{\lfloor w/2\rfloor}x_{t+i},\qquad t=\lfloor w/2\rfloor+1,\ldots,N-\lfloor w/2\rfloor\label{eq:1}
\end{equation}

where $x_{t}$ denotes the input time-series signal of length $N$,
and $w$ is the window size of the simple moving average (SMA) filter.
This centered formulation ensures that the smoothing operation does
not introduce temporal lag or phase shift. The valid index range for
$t$ guarantees that the averaging window lies entirely within the
signal boundaries, thereby avoiding boundary artifacts. No padding
or extrapolation is applied at the signal edges.

To evaluate the predictive performance of our proposed approach, we
established a benchmark suite comprising several machine learning
architectures, as shown in table \ref{tab:model_comparsion}. Specifically,
we developed and implemented: (i) a classic multilayer perceptron
(MLP); (ii) a long short-term memory (LSTM) network, representative
of widely used time-series models; (iii) a Transformer encoder based
model, which forms the basis for Google's BERT \citep{Devlin2018};
(iv) a Transformer decoder based model, as used in OpenAI's ChatGPT
\citep{OpenAI2024,Brown2020}; and (v) a Transformer-based model \citep{Vaswani2017}.

\begin{table}
\caption{\label{tab:model_comparsion} List of tested models with mean squared
error (MSE) loss on validation set. All models employ Bayesian optimization
to identify relatively optimal architectures. The losses reported
in the table correspond to those obtained under each model's optimal
configuration.}

\centering{}%
\begin{tabular}{cc}
\hline 
Model type &
Mean Squared Error (MSE) loss\tabularnewline
\hline 
Multilayer perceptron (MLP) &
0.0224\tabularnewline
Long Short-Term Memory (LSTM) &
0.015\tabularnewline
Transformer Encoder &
0.011\tabularnewline
Transformer Decoder &
0.011\tabularnewline
Our transformer-based model &
\textbf{0.0096}\tabularnewline
\hline 
\end{tabular}
\end{table}

All models employed Bayesian optimization \citep{Bergstra2011} to
determine relatively optimal architectural configurations. Each model
was trained and evaluated using the same 550 discharges. The loss
values reported in table \ref{tab:model_comparsion} correspond to
those obtained under each model's optimal setup. As shown in table
\ref{tab:model_comparsion}, our Transformer-based model achieves
the lowest mean squared error (MSE) loss among all tested models.
In machine learning evaluation, a smaller MSE loss indicates better
predictive performance. From a machine learning perspective, this
result is consistent with the ability of Transformer models to capture
long-range temporal dependencies and complex cross-channel interactions
through self-attention mechanisms, which are particularly well suited
to the multi-signal, long-duration discharge sequences considered
in this study. Based on both the superior performance shown in table
\ref{tab:model_comparsion} and these architectural advantages, the
Transformer-based model was selected for this work.

Our machine learning model was developed using PyTorch on Red Hat
Enterprise Linux 8, running on four A100 GPUs. During model training,
we utilized Bayesian algorithm \citep{Bergstra2011} to perform the
architectural hyperparameter search. Additionally, we experimented
with various optimizers and regularization techniques, ultimately
identifying the optimal set of hyperparameters, as shown in table
\ref{tab:hyperparameters}. Inference time was further tested on both
a consumer-grade RTX 3090 GPU and an A100 GPU, with both achieving
similar inference times of $\sim0.1$ seconds.

\begin{table}
\caption{\label{tab:hyperparameters}Our model hyperparameters}

\centering{}%
\begin{tabular}{ccc}
\toprule 
Hyperparameter &
Explanation &
Best value\tabularnewline
\midrule
$\text{\ensuremath{\eta}}_{t}$ &
Learning rate &
$1\text{\ensuremath{\times10^{-3}}}$\tabularnewline
num\_heads &
Number of heads &
8\tabularnewline
num\_layers &
Number of layers &
2\tabularnewline
Optimizer &
Optimizer type &
Stochastic Gradient Descent (SGD)\tabularnewline
$\eta_{\text{dropout}}$ &
Dropout rate &
0.1\tabularnewline
B &
Batch size &
16\tabularnewline
W &
Window size &
1024\tabularnewline
S &
Step size &
512\tabularnewline
Filter &
Filter type &
Simple moving average\tabularnewline
$W_{\text{filter}}$ &
Window size of the filter &
11\tabularnewline
\bottomrule
\end{tabular}
\end{table}

\section{Results \label{sec:Results}}

This section presents the results and analysis. The results include
different types of discharges and the statistical evaluation of six
output signals. In this study, we randomly partition the entire dataset
into training, validation, and test sets with a ratio of 6:2:2. The
machine learning model is trained on the training set, and the optimal
model is selected based on its performance on the validation set.
The results presented in this section correspond to the model's performance
on the test set, which includes 110 discharges. The final average
mean squared error (MSE) on the test set is 0.026, and the coefficient
of determination $R^{2}$ is 0.94.

\begin{figure}
\begin{centering}
\includegraphics[width=0.33\textwidth]{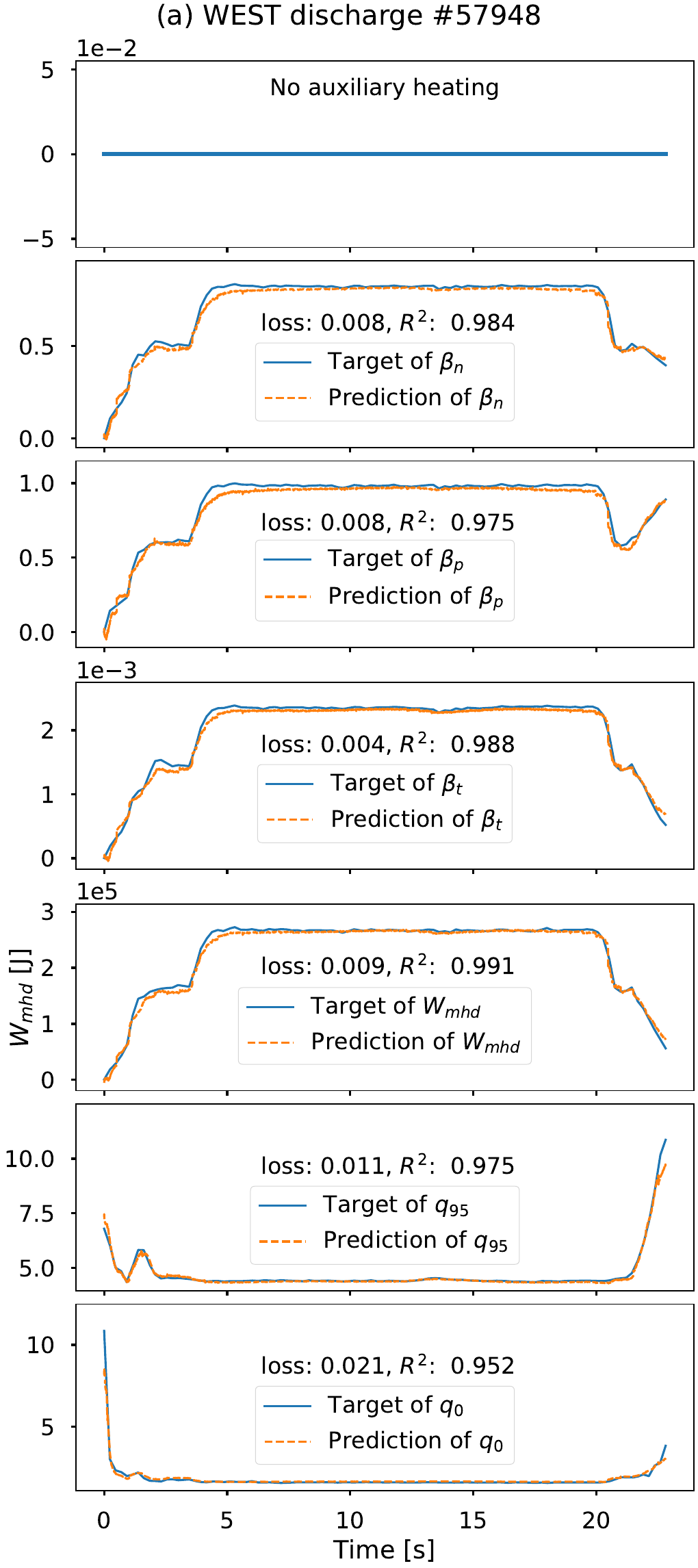}\includegraphics[width=0.33\textwidth]{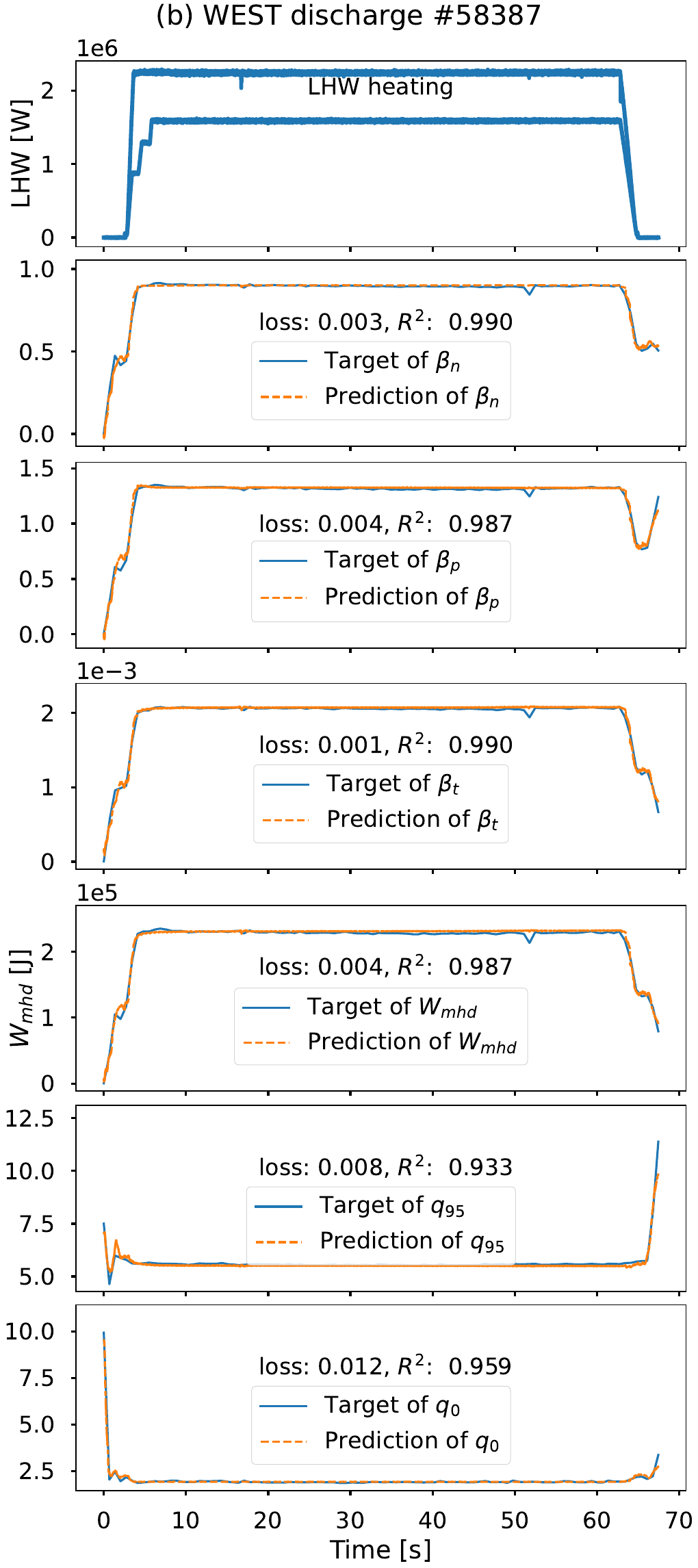}\includegraphics[width=0.33\textwidth]{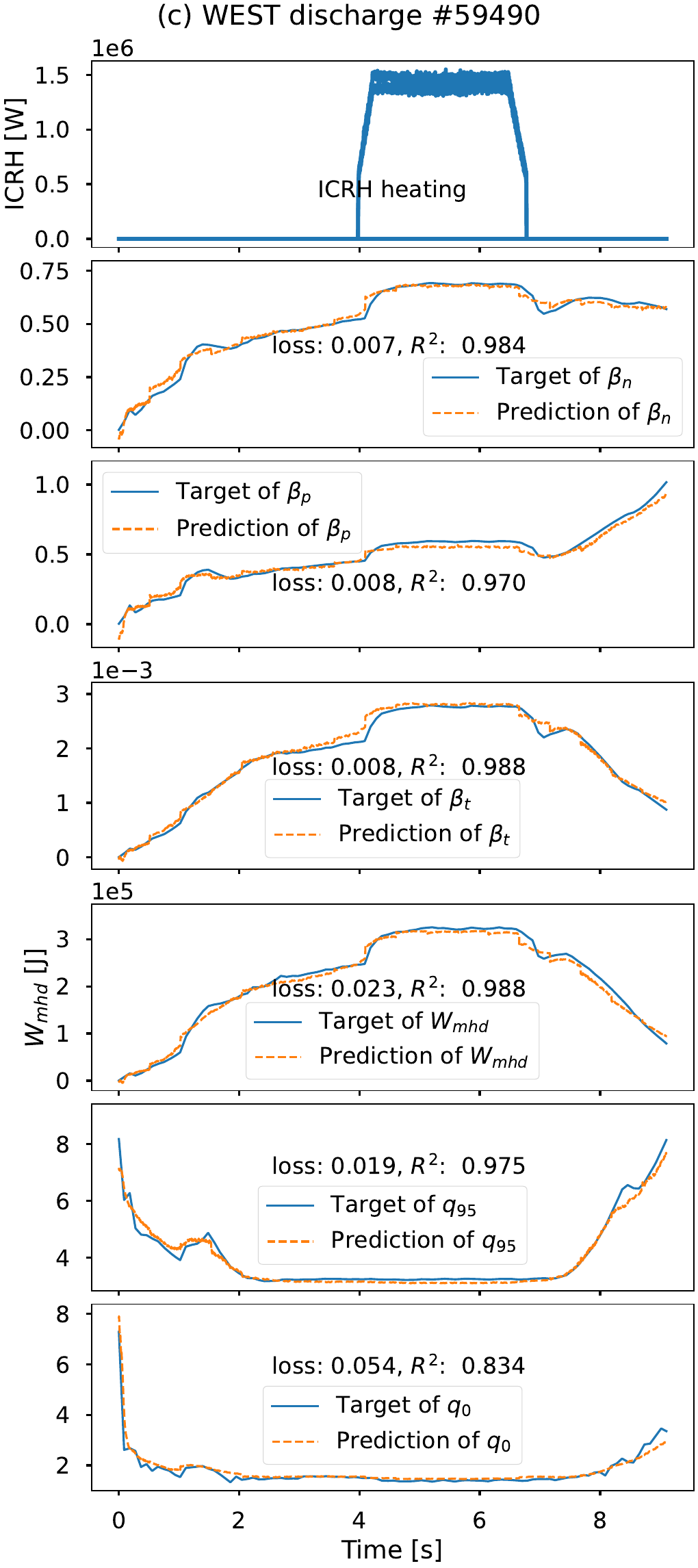}
\par\end{centering}
\caption{\label{fig:discharge_types} Three types of typical discharge prediction
in the test set. (a) The Ohmic heating-only discharge. (b) The discharge
with LHW heating. (c) The discharge with ICRH heating. The targets
represent experimental measurements, while the predictions correspond
to the model estimations.}
\end{figure}

\begin{figure}
\begin{centering}
\includegraphics[width=1\textwidth]{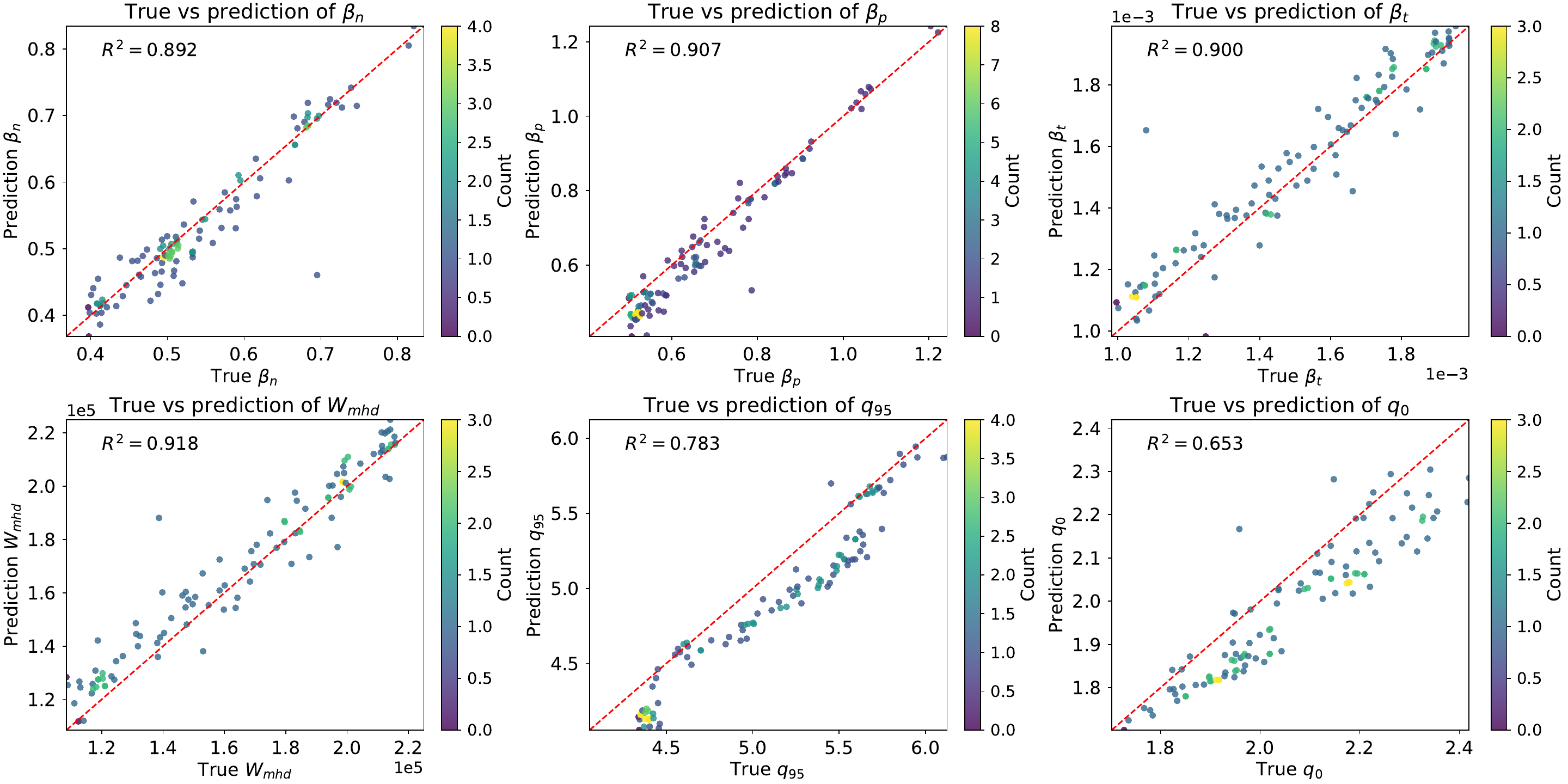}
\par\end{centering}
\caption{\label{fig:stat_true2pred} Regression plots of the output signals.
Both the target and predicted values are averaged over the duration
of each discharge. Except for $q_{0}$ and $q_{95}$, the model achieves
a coefficient of determination $R^{2}$ greater than 0.890. The performance
for $q_{0}$ and $q_{95}$ is lower compared to the other signals.}
\end{figure}

Figure \ref{fig:discharge_types} shows the model predictions for
three typical WEST discharges \#57948, \#58387, \#59490. The duration
of these discharges ranges from approximately 10 to 70 seconds, and
they involve 2 types of auxiliary heating. Under these conditions,
our model can successfully reproduce the entire discharge pulse from
the ramp-up, through the flat-top, to the ramp-down phases. However,
there is a slight discrepancy in the predictions of $q_{0}$ and $q_{95}$.
We also observe the same phenomenon in statistical plots as shown
in figure \ref{fig:stat_true2pred}. These quantities exhibit lower
accuracy because our input signals do not properly contain the pressure
profiles and kinetic measurements. Our goal is to reproduce the tokamak
response using signals that can be specified before discharge, these
measurements should not be included in the input signals. If $q_{0}$
and $q_{95}$ are not properly constrained, they may be unreliable
due to large variances, as the model struggles to reconstruct these
parameters. In addition, even under well-controlled experimental conditions
with stable wall behavior, identical or highly similar input configurations
can still lead to different plasma responses. This challenge is particularly
pronounced for $q_{0}$ and $q_{95}$, which depend on the radial
distribution of the plasma current density $j(r)$ \citep{Levinton1993}.
Since most of the input signals do not uniquely determine the internal
current redistribution during the discharge, these quantities are
weakly identifiable from pre-discharge inputs alone.

As shown in figure \ref{fig:stat_true2pred}, we quantitatively evaluate
the model performance on the entire test set, which consists of 110
discharges. For each discharge, the target and predicted values are
averaged over its duration. Overall, the model demonstrates strong
performance across most output parameters, except for $q_{0}$ and
$q_{95}$, where prediction accuracy is relatively lower. A small
number of discharges show poor predictive performance. Detailed analysis
reveals that these cases are mainly associated with fluctuations in
magnetic field coil currents, events that are uncommon and not part
of standard tokamak operational planning (e.g., discharge \#57821,
see Appendix A, figure \ref{fig:WEST-57821}). Such cases deviate
from our objective of reproducing plasma responses in conventional
discharges and providing reliable references for tokamak scenario
design, as these fluctuations are typically unintended and fall outside
the scope of standard design procedures. The systematic underestimation
of $q_{0}$ and $q_{95}$ does not originate from a structural limitation
of the machine learning model. Similar behavior is observed across
different architectures (e.g., in our LSTM-based model tests), indicating
that the effect is primarily data-driven. Preprocessing steps such
as signal smoothing and overlap-averaging can attenuate rapid temporal
variations, leading to reduced peak values near transient phases.
In addition, equilibrium reconstruction of $q$ profiles may introduce
asymmetric noise or offsets, which, under an MSE-based training objective,
can result in regression-to-the-mean behavior. Together, these factors
contribute to the observed underestimation.

In addition, some discharges correspond to rare operational scenarios,
such as discharge \#60147 (see Appendix A, figure \ref{fig:WEST-60147}),
which involves the use of two valid auxiliary heating methods and
they are very rare in our dataset. This discharge occurs only three
times out of 550 discharges. All of these cases are in the test set.
The model's generalization ability for such unusual and rare discharges
remains limited. Handling rare cases is an inherent challenge for
machine learning methods. To improve performance in future work, we
plan to include more discharges from this scenario to enhance the
model\textquoteright s exposure and learning. We will also explore
some data augmentation for the rare and fluctuating discharges, or
use some physics-informed method to enhance the ML model's generalization
ability.

\section{Conclusion and discussion \label{sec:Conclusion-and-discussion}}

In this study, we proposed a transformer-based machine learning model
for forecasting the behavior of tokamak plasmas based on discharge
configurations. The model takes as input the auxiliary heating power,
magnetic field coil currents, plasma current, and line-averaged plasma
density. These signals are typically designed or programmed prior
to the actual discharge, enabling the model to predict the plasma's
response to the intended operational settings.

The objective of the present work is complementary rather than competitive
with real-time physics-based simulation frameworks. By learning the
direct mapping between pre-discharge inputs and global plasma responses
from a large experimental dataset, the proposed model bypasses explicit
transport assumptions and empirical coefficient tuning. This enables
fast, fully pre-discharge prediction that reflects the actual behavior
of the machine. In this sense, physics-based reduced models answer
the question of what should happen according to simplified physical
assumptions, whereas the proposed data-driven approach captures what
actually happens, as evidenced by historical experimental operation.

The current model focuses on predicting six key global plasma parameters:
normalized beta ($\beta_{n}$), toroidal beta ($\beta_{t}$), poloidal
beta ($\beta_{p}$), plasma stored energy ($W_{\mathrm{mhd}}$), safety
factor at the magnetic axis ($q_{0}$), and safety factor at the 95\%
flux surface ($q_{95}$). These quantities serve as essential indicators
of plasma performance, stability, and confinement quality. The model
demonstrates promising predictive performance and offers potential
applications in discharge planning and fast scenario evaluation for
tokamak experiments.

The present work can be extended to predict additional plasma parameters,
such as the loop voltage ($V_{\mathrm{loop}}$) and internal inductance
($l_{i}$) or 1D profiles such as density and temperature, which are
also important for characterizing plasma behavior and optimizing control
strategies. However, a significant challenge lies in obtaining sufficient
and diverse data to properly constrain the model and ensure its generalization
to a broader range of operational scenarios.

Moreover, purely data-driven models exhibit strong sensitivity to
the specific conditions of the device. Consequently, directly applying
the model after major hardware modifications or transferring it to
a different tokamak remains an unresolved challenge. To address these
limitations, future research can explore hybrid approaches that integrate
physical knowledge or constraints into the model architecture or training
process. Incorporating physics-informed methods has the potential
to enhance the model's robustness and generalization capabilities,
particularly for extrapolating to unfamiliar operational regimes.

Additionally, collecting datasets from multiple devices and adopting
dimensionless representations for both input and output signals can
further improve the cross-device applicability and generalization
capability of the model. These approaches align with the broader goal
of developing reliable, data-driven surrogate models that can support
discharge design, scenario optimization, and real-time plasma control
across present and future tokamak devices.

\section*{Data availability statement}

The data supporting the findings of this study are owned by WEST/CEA,
and the code is jointly owned by WEST/CEA and NTU. Both are available
from the corresponding author upon reasonable request.

\ack{}{}

The computational work for this article was partially performed on
resources of the National Supercomputing Centre, Singapore (https://www.nscc.sg).

This research is supported by the National Research Foundation, Singapore,
the National Natural Science Foundation of China under Grant No. 12505267,
the China Postdoctoral Science Foundation under Grant Nos. 2023M743541
and 2023M733545, the Anhui Postdoctoral Scientific Research Program
Foundation under Grant No. 2024C997, the Science Foundation of Institute
of Plasma Physics, Chinese Academy of Sciences No. DSJJ-2025-11, and
WEST team (http://west.cea.fr/WESTteam) under the NTU/CEA SAFE (Singapore
Alliance with France for Fusion Energy) collaboration agreement.

\FloatBarrier

\section*{Appendix A. The parameter distributions in our \hypertarget{AppendixA}{}}

\begin{figure}
\begin{centering}
\includegraphics[width=0.9\textwidth]{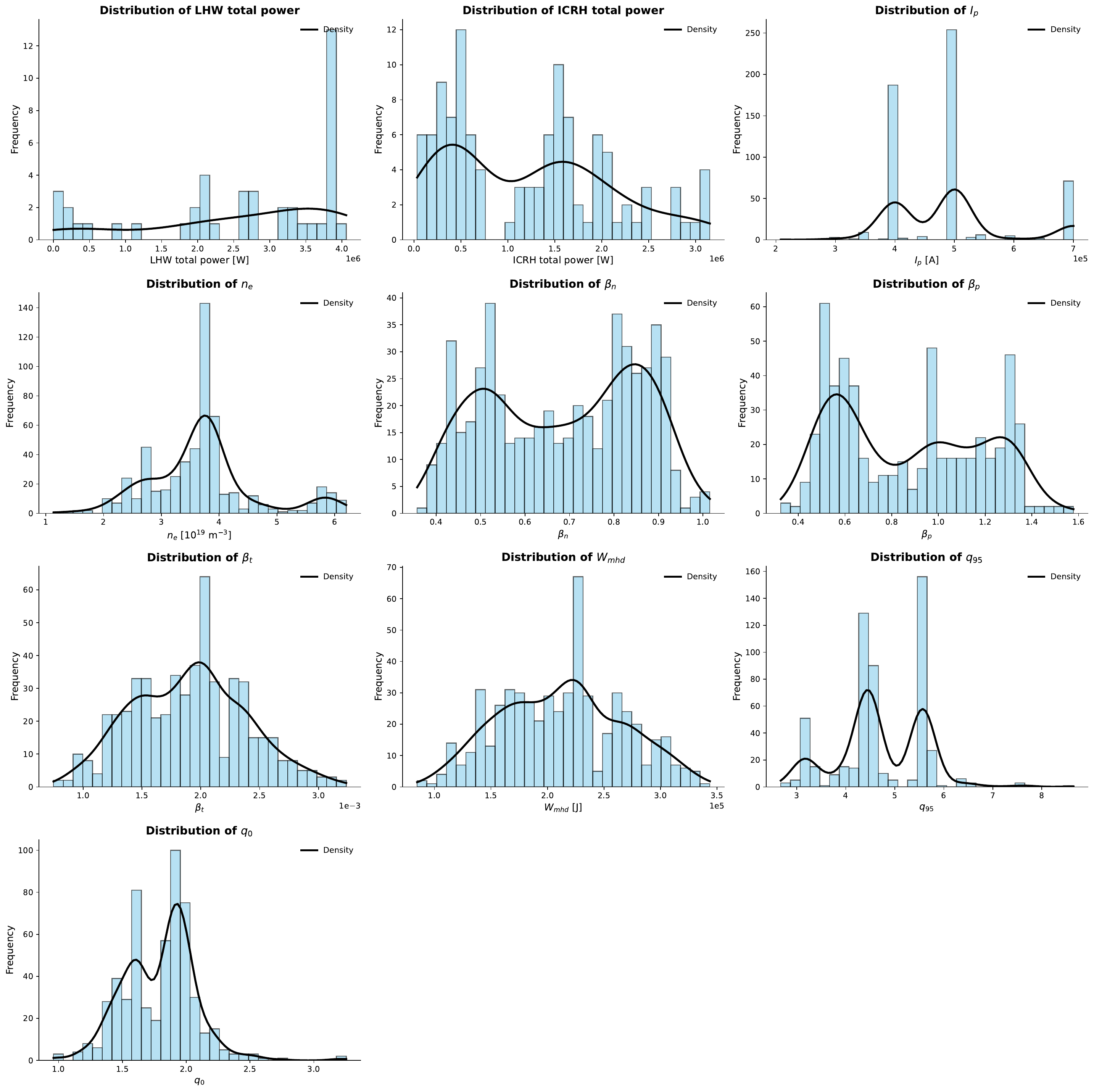}
\par\end{centering}
\caption{\label{fig:Output-dist} Discharge parameter distributions in the
whole dataset. All ICRH and LHW discharge with a total power of zero
are removed.}
\end{figure}

\FloatBarrier

\section*{Appendix B. Robustness of reconstruction with respect to sliding-window
step size \hypertarget{AppendixA}{}}

The final goal of the variance comparison is to understand the trend
of variance vs step size. To reduce computing complexity, the variances
in the figure are calculated as follows:

To characterize the trend of prediction stability with respect to
the sliding-window step size, we define an error-based variance metric
as 
\[
\mathrm{Var}(s)=\frac{1}{LK}\sum_{t=0}^{L}\sum_{k=1}^{K}\left(\hat{Y}_{t,k}(s)-Y_{t,k}\right)^{2},
\]

where $\hat{Y}_{t,k}(s)$ denotes the predicted value of the $k$-th
output feature at time step $t$ obtained with step size $s$, and
$Y_{t,k}$ is the corresponding ground-truth value. $L$ is the total
number of time samples aggregated over the evaluated discharge segments,
and $K$ is the number of output features, which is $K=6$ in this
study.

\begin{figure}
\begin{centering}
\includegraphics[width=0.6\textwidth]{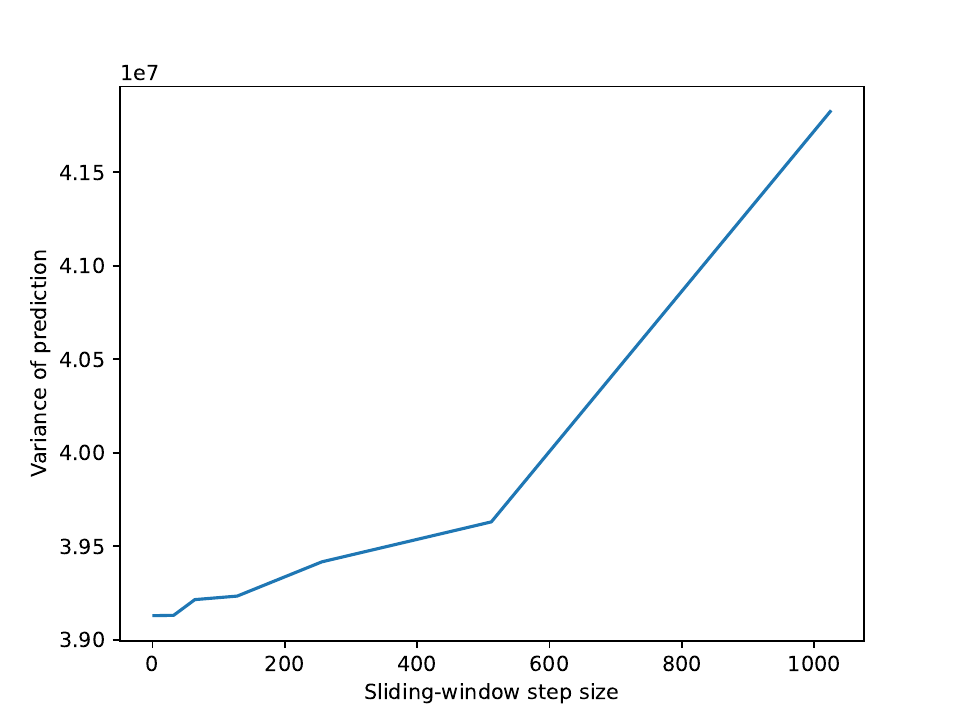}
\par\end{centering}
\caption{\label{fig:Variance} Variance of reconstructed predictions as a function
of sliding-window step size, evaluated using the same transformer-based
model as in the present work.}
\end{figure}

As shown in figure \ref{fig:Variance}, as the step size increases,
corresponding to reduced overlap between adjacent windows, the variance
of prediction increases, and vice versa. This occurs because partially
independent fluctuations across windows are suppressed. This trend
indicates that overlap-averaging enhances numerical stability and
robustness of the reconstruction.

\FloatBarrier

\section*{Appendix C. The typical bad case of our model. \label{sec:Appendix-C.}}

\begin{figure}[H]
\begin{centering}
\includegraphics[width=1\textwidth]{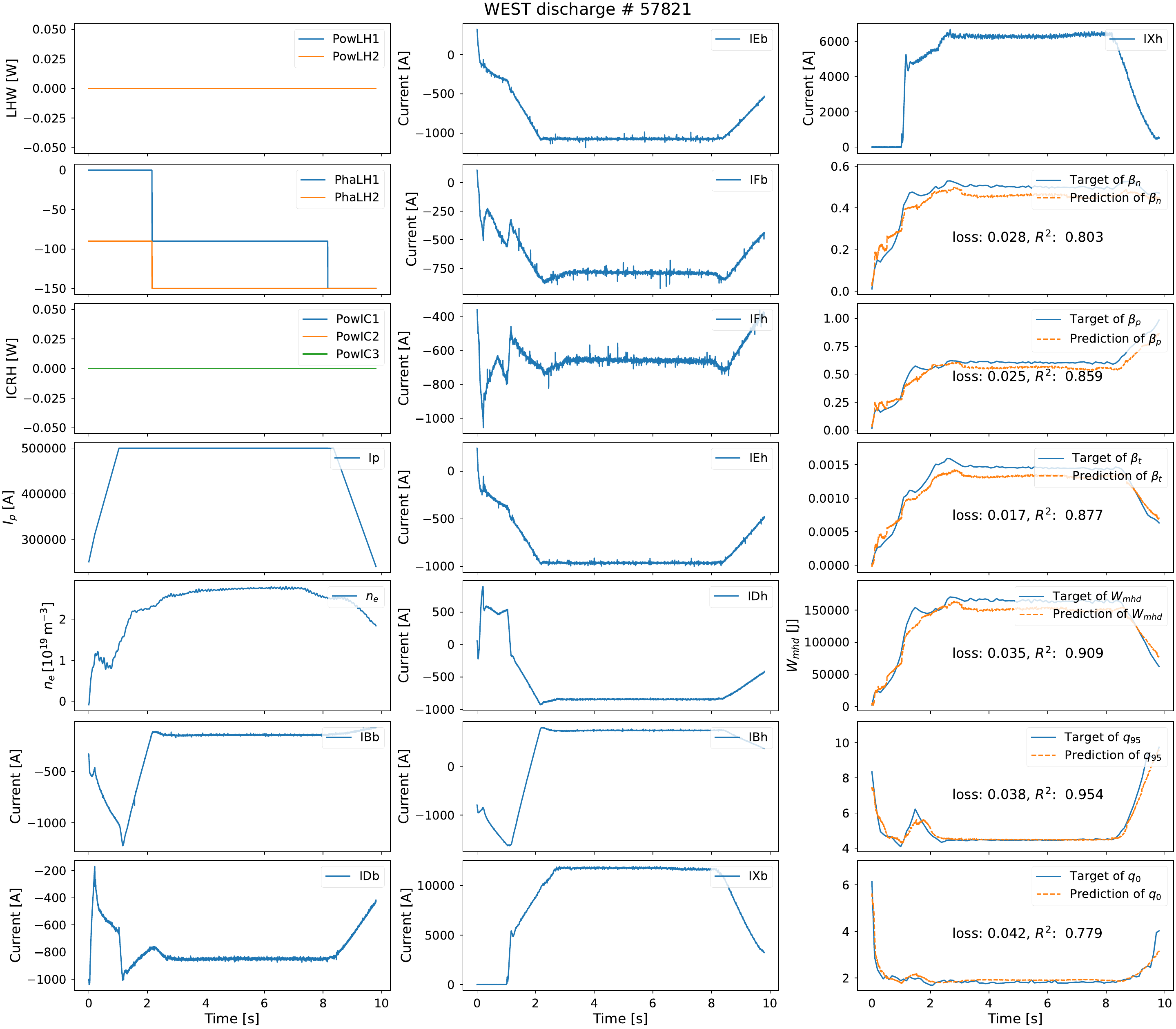}
\par\end{centering}
\caption{\label{fig:WEST-57821} WEST discharge \#57821. During the time interval
2\textendash 6 s, the PF coil currents exhibit a significant spike.
This irregularity in the input signals leads to discrepancies in the
model's output predictions.}
\end{figure}

\begin{figure}
\begin{centering}
\includegraphics[width=1\textwidth]{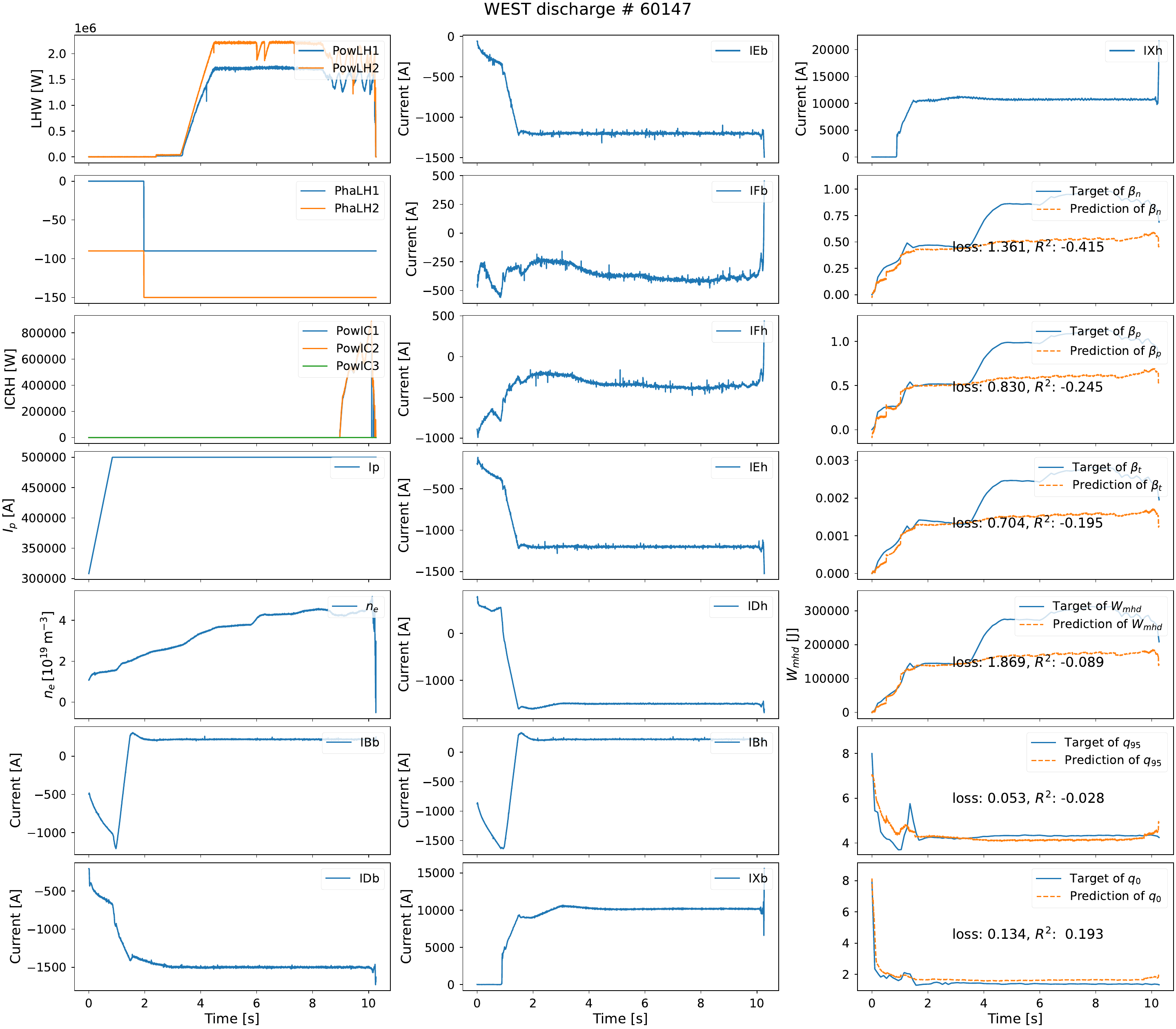}
\par\end{centering}
\caption{\label{fig:WEST-60147} WEST discharge \#60147. This type of discharge
uses two valid auxiliary heating methods, which is very rare in our
dataset. Only three out of 550 discharges exhibit such cases, all
of which are in the test set. As a result, the model has not learned
sufficient information to accurately predict these uncommon discharge
scenarios.}
\end{figure}

All signal names of figures \ref{fig:WEST-57821} and \ref{fig:WEST-60147}
follow the same notation as in figure \ref{fig:granger-pearson}

\FloatBarrier

\printbibliography

\end{document}